\documentstyle[epsfig,preprint,aps]{revtex}
\def\r#1{\ignorespaces $^{#1}$}         
\begin{document}
\draft
\onecolumn
\title{
\begin{center}
  Searching for the Higgs Boson at Hadron Colliders using the
Missing Mass Method
\end{center}
      }
\author{ Michael G. Albrow,\r {1} Andrey Rostovtsev\r {2}
}
\address{
\begin{center}
\r 1  { Fermi National Accelerator Laboratory, Batavia, IL 60510}\\
\r 2  { Inst. Theor. Exp. Phys. (ITEP), Moscow, Russia}\\
\end{center}
}
\maketitle

\begin{abstract}
{
If the Higgs is produced with a large enough cross section
in the {\em exclusive} reaction $p + \bar{p} \rightarrow p + H + \bar{p}$
it will give rise to a peak at $M_H$
in the {\em missing mass} ($MM$) spectrum, calculated from
the 4-momenta of the beam particles and
the outgoing $p$ and $\bar{p}$. 
The resolution
in $MM$ can be approximately 250 MeV, independent of $M_H$ from
100 GeV to 200 GeV. This high resolution makes a search feasible
over nearly this full mass range at the Tevatron with 15 fb$^{-1}$
as hoped for in Run II. 
}
\end{abstract}

%
%
\par
The predominant mode for Higgs production at hadron colliders is 
$gg$-fusion\cite{daws,spir} through a virtual
top quark loop. The dominant decay mode 
up to 135 GeV is to $b\bar{b}$, above which the $W W^*$ mode becomes increasingly
important until $M_H > 2M_W$ (160 GeV) when both $W$ are real.
By 200 GeV the $ZZ$ mode has grown to 26\%.
The $\tau^+\tau^-$ mode decreases from 7.6\% at 110 GeV to about 2\% at
150 GeV.
The intrinsic 
width of a Higgs over this mass region rises, from 5 MeV
at $M_H$ = 130 GeV, to 16 MeV at $M_H$ = 150 GeV,
to 650 MeV at $M_H$ = 180 GeV\cite{daws}, so
mass resolution is crucial in increasing the signal:background $S:B$
ratio. 

One has until now considered the observation of the Higgs 
in the intermediate mass region 110 GeV to 130 GeV in inclusive reactions
to be impossible because of the small $S:B$.
The mass resolution in reconstructing a $b\bar{b}$ di-jet
is about 10 GeV - 15 GeV, and the QCD background is indeed overwhelming when
the signal is so spread out. A high price has to be paid
to improve the $S:B$ ratio by selecting relatively rare cases where
it is produced in association with a massive particle ($W,Z,t$)
or where it decays to $\gamma\gamma$ (branching fraction $\approx 2 \times 
10^{-3}$), where much better mass resolution can be obtained than for the 
$b\bar{b}$ di-jets.

In the exclusive process $p + \bar{p} \rightarrow p + H + \bar{p}$, 
with no other particles in the final state (we talk in this 
note in Tevatron terms although all the arguments clearly refer
also to the LHC), we use the known 4-momenta of the incoming and
outgoing $p$ and $\bar{p}$ to calculate the missing mass from
$MM^2 = (p_{b1} + p_{b2} - p_3 - p_4)^2 $. The visibility of a
signal will depend on the spread in these quantities; any overall scale factor
such as would come e.g. from uncertainty in the magnetic fields in the Tevatron
only affects the central value, i.e. $M_H$ if a signal is seen.
The momentum spread of the incoming beams \cite{bagl}
is 1.0 $\times 10^{-4}$ at the beginning of
a store and rises to about 1.6 $\times 10^{-4}$ after 20 hours of collisions.
The position of the interaction point $x_\circ ,y_\circ ,z_\circ$ can be 
reconstructed with $\sigma \approx 4\mu$m, 4$\mu$m and 10$\mu$m
respectively for central $b\bar{b}$ jets, and about a factor two worse
\footnote{We assume both leptons are tracked in the silicon vertex detectors.}
for $l^+l^-$ final states. 
 The outgoing $p$ and $\bar{p}$ tracks can be measured after 
18.8m of 4.34 Tesla dipoles
using several layers of
crossed and tilted silicon pixel detectors giving $\sigma_x = \sigma_y
\approx 2.5 \mu$m over $\approx$ 1.0m, thus 
$\sigma_{x'} = \sigma_{y'} = 2.5 \times 10^{-6}$.
If $\sqrt{s}$ is the center of mass energy, 2 TeV at the Tevatron in Run II,
and the outgoing scattered beam particles have lost fractions $\xi_1, \xi_2$
of their incident momenta ($\xi = 1 - x_F$ where $x_F$ is Feynman-$x$),
we have approximately $MM^2 = \xi_1 \xi_2 s$.
The spread in the reconstructed missing mass, $\delta_{MM}$ 
is a combination of the relative spread $\frac{\delta p_b}{p_b}$
in the
beam particles' momenta $p_b$ and the resolution of the ``dipole spectrometers"
which use the primary interaction point and the outgoing track.
With the above parameters this is $\approx$ 250 MeV, independent of
$MM$. 

We note that this method is not limited to Higgs searches but
would be sensitive to any relatively narrow 
massive objects with vacuum quantum numbers.
\par
The visibility of the Higgs by this technique
clearly depends on 
the size of the cross section
for the process where the Higgs is produced (in the central region)
completely exclusively, i.e. the $p$ and $\bar{p}$ go down the beam pipes
each having lost about $\frac{M_H}{2}$ in longitudinal momentum
and no other particles are produced. The
mechanism is as usual $gg \rightarrow H$ through intermediate loops of
heavy particles (predominantly a top loop); this normally leaves the
$p$ and $\bar{p}$ in color-octet states and gives rise to 
color strings filling rapidity with hadrons. However some fraction of the time
one or more other gluons can be exchanged which neutralize 
(in a color sense) the 
$p$ and $\bar{p}$ and can even leave them in their ground state.
In Regge theory this is the double pomeron exchange
($DPE$) process. Several attempts have been made to calculate this cross
section. 
In 1990 Sch\"{a}fer, Nachtmann and Sch\"{o}pf\cite{scha}      
considered diffractive Higgs production at the LHC and SSC,
concluding that the cross sections for the exclusive process
could not be reliably predicted. 
M\"{u}ller and Schramm\cite{mull} made a calculation, also for
nucleus-nucleus collisions, and concluded that the exclusive
process is immeasurably small.
In 1991 Bialas and Landshoff\cite{bial}
calculated from Regge theory that about 1\% of all Higgs events may
have the $p$ and $\bar{p}$ in the $DPE$ region of $x_F \approx$ 0.95,
but they did not estimate the {\em fully exclusive} cross section.
In 1994 Lu and Milana\cite{lumi} obtained an estimate ``well below what is
likely to be experimentally feasible".
In 1995 Cudell and Hernandez\cite{cude} made a lowest order QCD calculation
with the non-perturbative form factors of the proton tuned
to reproduce elastic and soft diffractive cross section measurements.
They presented the exclusive production cross section as a function
of $M_H$ up to 150 GeV at $\sqrt{s}$ = 1.8 TeV.
They found a cross section decreasing slowly 
with $M_H$ from 45 fb at 110 GeV, 13.5 fb at 150 GeV and, by extrapolation,
6.0 fb at 170 GeV (all within a factor two).
The total Higgs production cross section by the dominant
$gg$-fusion mechanism is \cite{spir} 900 fb, 364 fb and 247 fb respectively so
the exclusive fraction decreases from 5\% to about 2.4\% over this mass
range, even higher than the Bialas and Landshoff estimate. 
There are two very recent
calculations. Khoze, Martin and Ryskin\cite{khoz} find
$\sigma(p+p \rightarrow p+H+p)$ = 0.06 fb for $M_H$ = 120 GeV at $\sqrt{s}$
= 2 TeV
if the probability $S_{spect}^2$ not to have extra rescattering in the 
interaction is $S_{spect}^2$ = 0.1. 
Kharzeev and Levin\cite{khar} find much more optimistically 19 - 140 fb
for $M_H$ = 100 GeV at the Tevatron, but do not present the $M_H$-dependence.
Although there are serious differences in the theoretical predictions,
we shall show that the more optimistic predictions allow a Higgs discovery
at the Tevatron in Run II over the full mass range from 110 GeV to 180 GeV.
We take the Cudell and Hernandez ($CH$) prediction as our benchmark,
ignoring any gain from the $\sqrt{s}$ increase from 1.8 TeV to 2.0 TeV
and noting that the $CH$ estimate has a factor $\approx$ 2 uncertainty.
It will be seen that even if the true exclusive cross section is lower by an
order of magnitude a discovery is still possible over most of this mass range. 
\par
We now consider signals and backgrounds, first for $b\bar{b}$, then for $\tau^+\tau^-$
and lastly for $WW^{(*)}$.
Table 1 shows a compilation of results. 

\begin{table}
\begin{center}
\begin{tabular}{lcccccc}
\hline
$M_H$  & $\sigma(CH)$ & Mode &  BR & $\sigma$.BR.BR & Events & Background\\
(GeV)  & (fb)         &      &     & (fb) &15fb$^{-1}$& /250 MeV\\
\hline
 110        & 45               & $b\bar{b}$     & 0.770 & 34.6 &260 &3.75\\
            &                  & $\tau^+\tau^-$ & 0.076 & 3.4 & 26& $<$ 0.1\\
\hline 
 130        & 25               & $b\bar{b}$     & 0.525 & 13.1 & 96&0.75\\
            &                  & $\tau^+\tau^-$ & 0.054 & 1.35 & 10.0 & $<0.1$ \\
            &                  & $WW^*$         & 0.289 & 0.72 & 5.4 & $\ll 1$\\
\hline
 150        & 13.5             & $WW^*$         & 0.685 & 0.93 & 7.0& $\ll 1$\\
\hline
 170        & 6.0              & $W^+W^-$       & 0.996 & 0.58 & 4.3 &$\ll 1$\\
\hline
 180        & 3.5              & $W^+W^-$       & 0.935 & 0.34 & 2.5& $\ll 1$\\
\hline
\end{tabular}
\end{center}

\caption{For various Higgs masses, the exclusive 
production cross section
according to Cudell and 
Hernandez at 1.8 TeV.
Column 5 shows the cross section $\times$ branching fractions
either to two b-jets or to two charged leptons.
A factor 0.5 has been applied to events and background for
acceptance/efficiency.}

\end{table}

For the $b\bar{b}$ dijet background we take CDF's published cross 
section\cite{cdfb} $\frac{d\sigma}{dM_{JJ}}$
for 
two b-tagged jets, which starts at 150 GeV, and extrapolate the fit to
the data (which is a factor 2-3 higher than the PYTHIA prediction) down to 
110(130) GeV finding 200(40) pb/GeV (in $|\eta| < 2.0, |cos(\theta^*)| < 2/3$).
From our other $DPE$ studies, of lower mass dijets\cite{cdpe}, we expect that
about $10^{-5}$ of these are 
DPE ($p + \bar{p} \rightarrow p +G+ b + \bar{b} +G+ \bar{p}$),
where $G$ represents a rapidity gap exceeding about 3 units,
assuming this fraction is not $E_T$-dependent. If the fraction is
smaller, so much the better. That gives 
0.5(0.1) fb per 250 MeV bin, 
to be compared with a signal of around 45(25) fb \cite{cude}.
With 15 fb$^{-1}$ and assuming 50\% acceptance 
for both signal and background we have 260(96) events 
(see Table 1) on a 
background of 3.75(0.75). Even if the $CH$ predictions are optimistic
by an order of magnitude 
these signals exceed 10$\sigma$. 
We have not put in a factor for b-tagging efficiency
(which affects the signal and the background the same way
apart from differences in the angular distributions); in CDF it was 
about 35\% per jet in Run I at $M_{JJ} = 200$ GeV. It will be higher 
in Run II with more silicon coverage and at
smaller masses; also we only have to tag one jet, so this is probably 
a very modest reduction in both signal and background. We have put in 
an acceptance of 50\% for the forward $p$ and $\bar{p}$ for the signal
and background, 
assuming the 
$|t|$-distribution is as expected for high mass $DPE$. 
The $S:B$ ratio rises with $M_H$ in this mass region 110-130 GeV.
\par
The Higgs branching fraction to $\tau^+\tau^-$ drops from 7.6\% at 110 GeV
to 5.4\% at 130 GeV, as the $WW^*$ mode grows in competition.
Backgrounds to the proposed search could come from normal Drell-Yan ($DY$)/$Z$
production together with 0,1, or 2 associated high-$x_F$ tracks; in the first 
two cases leading (anti-)protons come from different events (pile-up);
we discuss ways of minimizing this later. In the third case the events
look like continuum $DPE$ production of $DY$ pairs, together with
associated particles. CDF found\cite{cdfw} single diffractive ($SD$)
production of $W$ at the level of (1.15 $\pm$ 0.55)\%
of non-diffractive ($ND$) production. A recent CDF study\cite{cdpe} of 
jet production at low $E_T$ has found 
a breakdown of factorization for jet production in the sense that 
$\frac{\sigma_{DPE}}{\sigma_{SD}} \approx 5 \times \frac{\sigma_{SD}}{\sigma_{ND}}$. 
Let us assume this fraction is the same
for high-mass $DY$, and then assume (conservatively) factorization 
break-down by the same factor 5 for high mass $DY$. Then
$DPE$ production of high mass $DY$ is at the relative level of 5.$10^{-4}$.
From a CDF study\cite{abdy} of high mass $e^+e^-$
and $\mu^+\mu^-$ we infer that $\frac{d\sigma}{dM}$ for the region
110-130 GeV is $100\pm 40$ fb GeV$^{-1}$. Therefore the cross section 
for $p \bar{p} \rightarrow p G \mu^+ \mu^- X G \bar{p}$, where $X$ represents
additional associated hadrons, $n_{ass}$ of which are charged tracks,
is expected to be about 100 fb GeV$^{-1}$ $\times$ 5.10$^{-4}$ =
0.05 fb.GeV$^{-1}$ or 0.2 events in 15 fb$^{-1}$ in a 250 MeV bin.
Note however that for the exclusive Higgs production process $n_{ass}$ = 0,
while for generic $DY/Z$ production $<n_{ass}> \approx 16$\cite{cdrf}
for $p_T \ge 0.2$ GeV, $|\eta| \le$ 1. We claim that the observation of
lepton pairs with no associated tracks, $n_{ass}$ = 0, would already
be good evidence for exclusive Higgs production 
The $CH$ cross section
$\sigma(p + \bar{p} \rightarrow p + H + \bar{p})$ $\times$ branching fraction
$H \rightarrow \tau^+\tau^-$ of 3.4 (1.3) fb 
at 110 (130) GeV gives 26 (10) events on a background of
less than 1 event if we include a 50\% acceptance/efficiency factor. 
High $p_T$ $\tau$ are easily recognized:
one-prong decays are 85\% and three-prong are 
15\%. A high $p_T$ 3-prong $\tau$ decay is quite distinct 
from a QCD hadronic jet
because it is tightly collimated, 
with $M_{eff} < M_{\tau}$ = 1.78 GeV.
\par
The Higgs branching fraction to $WW^{(*)}$ rises from 29\% at 130 GeV
to 69\% (97\%) at 150 (170) GeV (see Table 1). Beyond 180 GeV it falls because
of competition from the $ZZ^{(*)}$ mode. We will only consider the
leptonic decay modes of the $W$ because of the spectacular cleanliness
of the event vertices: either $ee,e\mu, \mu\mu, e\tau, \mu\tau$
or $\tau\tau$ and no other charged particle tracks ($n_{ass}$ = 0), 
together with large
$E\!\!\!\!/_T$ and the forward $p$ and $\bar{p}$.

Precision timing 
($\approx$ 30 ps) 
on the $p$ and $\bar{p}$ will not only check that they came from
the same interaction but
can pin down the vertex $z_{vtx}$ to about 1 cm.
To estimate the signal we extrapolate the Cudell and Hernandez (1.8 TeV) 
exclusive
cross sections from 150 GeV (11 - 16 fb) to 180 GeV (2.5 - 5 fb).
Putting in $BR(H \rightarrow W W^{(*)})$, a 10\% probability that both
$W$ decay leptonically, and assuming that, by using lower than usual
trigger thresholds on the central leptons and $E\!\!\!\!/_T$, we can keep the
efficiency at 50\%, we find in 15 fb$^{-1}$ 7 events for
$M_H$ = 150 GeV falling to 2.5 events at $M_H$ = 180 GeV.
To estimate the background we refer to the observation of five
$W^+W^-$ events by CDF\cite{cdww} 
\footnote{D\O\  earlier found one $e^+e^-$ event\cite{abac} in 14 pb$^{-1}$.}
which gave 
$\sigma(p + \bar{p} \rightarrow W^+W^- X)$ 
= 10.2 $\pm$ 6.5 pb which we assume to be roughly
uniform over $160 < M_{WW} < 180$ GeV
so $d\sigma/dM \approx$ 0.5 pb GeV$^{-1}$. Below 160 GeV the cross section
for $WW^*$ will be smaller. 
The observed $W^+W^-$ cross sections are consistent with
Standard Model NLO expectations, ignoring the Higgs, of 
$\sigma(p + \bar{p} \rightarrow W^+W^- X)$ = 10 pb at 1.8 TeV. 
We multiply by the 10\% probability that {\em both} $W$
decay leptonicaly and apply a
50\% ``efficiency" for detecting the $p,\bar{p}$ and both leptons
and recognizing the event as $l^+l^- E\!\!\!\!/_T$. This is high compared with
the efficiency in ref \cite{abew}, which was 5.4\% - 8.9\%, because due to the
lack of background we can surely lower the selection cuts on $E\!\!\!\!/_T, p_T(e),
p_T(\mu)$ and $p_T(\tau)$ significantly. 
We assume as before that
about 5 $\times$ 10$^{-4}$ of these are from $DPE$,
giving $\approx 3 \times 10^{-3}$ fb/250 MeV.
For any non-diffractive background we can asssume that
the associated charged multiplicity on the $WW$ vertex is Poisson-distributed 
with a
mean of about 16, which is what CDF observes\cite{cdrf} for $Z$ events. This non-diffractive
background then has a completely negligible tail at $n_{ass}$ = 0.
Thus the backgrounds in all the dilepton channels with $n_{ass}$ = 0
are negligible, and even 3 or 4 events at the same $MM$ 
would constitute a discovery. 
\par
In order not to be limited by the number of interactions in a 
bunch crossing one should not use a method requiring rapidity gaps
(as normally measured in counters or calorimeters). This is where the strength 
of using only leptonic decays of the $W^+W^-$ enters.
Tracking back the $l^+$ and $l^-$
to their common vertex (which can be done using the SVX detectors
in CDF to a precision $\sigma_x = \sigma_y \approx 10 \mu$m 
and $\sigma_z < 20 \mu$m )
there will, for the exclusive process, be {\em no other particles} coming
from the same vertex, $n_{ass}$ = 0.
All ``normal" production of $W$-pairs will on the contrary
have a highly active vertex with many associated hadrons.
(Even in the absence of dipole spectrometers one can plot
$n_{ass}$ and look for a peak at $n_{ass}$ = 0. This would be
``evidence" for exclusive Higgs production. Enough events of this
kind would enable one to make fits of kinematic quantities
as a function of $M_H$.)
One can then plot the missing mass $MM$ for these superclean events with
two and only two oppositely charged leptons on a vertex, with and
without $E\!\!\!\!/_T$.
A Higgs signal will be a cluster of events at the same $MM$ within the
resolution ($\approx 250$ MeV).

If the exclusive cross section is indeed big enough to
provide a few events in the data, but continuum background were to be an issue,
one has further recourse to angular distributions\cite{ditt}. 
The $H$ is a scalar and
decays isotropically, while generic $W^+W^-$ production is not
isotropic with respect to the beam axis; also
the $W$'s (like the $\tau$'s) will have opposite polarizations. This is not
generally true for the backgrounds, so one can plot quantities sensitive
to these kinematic features as a function of $MM$ to look for 
localised structure.
\par
\par 
With multiple interactions in a bunch crossing 
a background could come from two single diffractive
collisions, one producing the $p$ and the other the $\bar{p}$.
One way of reducing this is to require longitudinal momentum balance.
However ``pile-up"
can be reduced by one to two orders of magnitude 
by backing up the silicon detectors
in the pots by counters with excellent timing resolution. 
A conventional fast detector would be a quartz (for radiation hardness)
block producing Cerenkov light viewed by a fast photomultiplier.
One can achieve $30$ ps timing resolution on the
$p$ and $\bar{p}$, much better than the ($\approx$ 1 ns) spread between random
concidences. There are ideas\cite{bros} for Fast Timing Cerenkov Detectors
($FTCD$) using microchannel plates which might achieve a resolution
of a few ps.
The sum of the $p$ and $\bar{p}$ times is a constant
for genuine coincidences, and their difference $\Delta t$ is a measure
of $z_{vtx}$ at the level of 1 cm
(for $\Delta t$ = 30 ps). 
\par
In Run IIA both CDF and D\O\ will have one dipole spectrometer arm.  
There are various studies that can be done already in Run IIA, before
the second arm spectrometer is installed, to learn more about 
the feasibility of this proposed Higgs search. 
\par
1) Measure the $b\bar{b}$ dijet mass spectrum, $M_{b\bar{b}}$, over
the mass range up to 150 GeV to complement the earlier CDF
measurement\cite{cdfb}. 
\par
2) Measure the $l^+l^-$
mass spectrum in the region
of $M_{l^+l^-}$ 80-180 GeV, carefully studying the
associated charged multiplicity $n_{ass}$ on the primary $l^+l^-$ vertex
for different mass ranges. 
\par
3) Measure the production of exclusive $\chi^\circ_c$ and 
$\chi^\circ_b$ states. 
Note that some of these states have the same quantum numbers
\footnote{Allowed quantum numbers for exclusive states 
in $DPE$ are $I^C =0^+$
but any $J^P$\cite{pump}.},
$I^GJ^{PC}
=0^+0^{++}$, as the vacuum and the Higgs. 
\par
In summary, using the missing mass method that
we propose, the resolution in Higgs mass can be
improved to 250 MeV, increasing the $S:B$ by a
factor $\approx$ 40 - 60.
The method works not only for $b\bar{b}$ Higgs decay but also for
$\tau^+\tau^-$, $W^+W^-$ and $ZZ$ decays, and the number of neutrinos 
in the final state is irrelevant for the mass resolution.
\par
This work was supported by the U.S. Department of Energy and the
Institute for Theoretical and Experimental Physics (ITEP),Russia. 
We thank P.Bagley and C. Moore for
information on the Tevatron, and V.Kim,  
D.Kharzeev and E.Levin for discussions on exclusive
Higgs production.

\end{document}